\documentclass[conference]{IEEEtran}
\usepackage{amsmath,amssymb,amsfonts}
\usepackage{algorithmic}
\usepackage{graphicx}
\usepackage{textcomp}
\usepackage{xcolor}
\def\BibTeX{{\rm B\kern-.05em{\sc i\kern-.025em b}\kern-.08em
    T\kern-.1667em\lower.7ex\hbox{E}\kern-.125emX}}
\usepackage{hyperref}
\usepackage{todonotes}
\usepackage{lipsum}
\usepackage{csquotes}
\usepackage{balance}
\usepackage{enumitem}

\usepackage{fancyhdr} % For custom headers and footers

\fancypagestyle{firstpage}{ 
  \fancyhf{} % Clear all header and footer fields for this specific page style
  \fancyhead[L]{%
    \textbf{\copyright2025 IEEE. Accepted for publication in: 2025 IEEE 33nd International Requirements Engineering Conference (RE).
Personal use of this material is permitted. Permission from IEEE must be
obtained for all other uses, in any current or future media, including
reprinting/republishing this material for advertising or promotional purposes, creating new
collective works, for resale or redistribution to servers or lists, or reuse of any copyrighted
component of this work in other works.}
  }
}

\newcommand{\textcite}[1]{\cite{#1}}
\begin{document}
\title{\vspace{40px}Human-Machine Collaboration and Ethical Considerations in Adaptive Cyber-Physical Systems}
% \\
% {\footnotesize \textsuperscript{*}Note: Sub-titles are not captured in Xplore and 
% should not be used}
% \thanks{Identify applicable funding agency here. If none, delete this.}
% }

\newcommand{\zp}[1]{\textcolor{purple}{#1}}

% I think it is a good proposal -- but I would make sure to anchor the research questions in 'what exists already?'  The one strange thing is the focus on 'novel' without asking what exists.  Also is it so important that they are novel?  Why not 'effective' instead?
		
% das hat jane geschrieben - das ist ein guter kommentar. es muss nicht quasi immer neu sein nur um des Neu-seins.
		
% Aber das machst du ja eigentlich eh, mit dem SLR, was gibt es schon, und wie kann man das effektiv einsetzen bzw. adaptieren.
		
% Das kann man da vielleicht genau so sagen 

% Absolut, das ist ein guter Punkt. ich denke in Section 3 kommt das aufbauende schon etwas raus. Im Falle kann ich dort auch noch mehr das Aufbauende hervorheben (bzw. kommt das dann auch mit der SLR). 

% > Aber das machst du ja eigentlich eh, mit dem SLR, was gibt es schon, und wie kann man das effektiv einsetzen bzw. adaptieren.

% So kann man auf jeden Fall argumentieren, dann ists 100% klar. Vermutlich wäre es Vorteilhaft, das schon in der Intro bei den Research Objectives zu erwähnen

\author{\IEEEauthorblockN{Zoe Pfister}
\IEEEauthorblockA{\textit{Deparment of Computer Science} \\
\textit{University of Innsbruck}\\
Innsbruck, Austria \\
zoe.pfister@uibk.ac.at}
}

% \IEEEoverridecommandlockouts
\maketitle
\thispagestyle{firstpage}

% \copyright2025 IEEE — Accepted for publication in: 2025 IEEE 33nd International Requirements Engineering Conference (RE) \hfill

\begin{abstract}
% Context and Motivation
Adaptive Cyber-Physical Systems (CPS) are systems that integrate both physical and computational capabilities, which can adjust in response to changing parameters.
Furthermore, they increasingly incorporate human-machine collaboration, allowing them to benefit from the individual strengths of humans and machines. 
Human-Machine Teaming (HMT) represents the most advanced paradigm of human-machine collaboration, envisioning seamless teamwork between humans and machines.
% Question / Problem (efficient, seamless, privacy-preserving integration)
However, achieving effective and seamless HMT in adaptive CPS is challenging.
While adaptive CPS already benefit from feedback loops such as MAPE-K, there is still a gap in integrating humans into these feedback loops due to different operational cadences of humans and machines.
Further, HMT requires constant monitoring of human operators, collecting potentially sensitive information about their actions and behavior. 
Respecting the privacy and human values of the actors of the CPS is crucial for the success of human-machine teams.
This research addresses these challenges by: (1) developing novel methods and processes for integrating HMT into adaptive CPS, focusing on human-machine interaction principles and their incorporation into adaptive feedback loops found in CPS, and (2) creating frameworks for integrating, verifying, and validating ethics and human values throughout the system lifecycle, starting from requirements engineering.
\end{abstract}

% The goal of this thesis is to advance research in the aforementioned areas, namely human-machine collaboration in CPS, mediator architectures in CPS, and integration of ethics in requirements engineering.
% More specifically, we aim to develop novel methods that allow for better collaboration between humans and machines, with current developments of HMT~\cite{cleland-huang_human-machine_2023} as a starting point. 
% Further, we will investigate how mediator architectures can be effectively introduced into existing CPS with their own respective paradigms, such as MAPE-K feedback loops~\cite{chambers_self-adaptation_2024}.
% Lastly, we will create methods and frameworks to integrate ethical considerations into the process of Requirements Engineering.

\begin{IEEEkeywords}
Value-Based Engineering, Adaptive CPS, Human-Machine Teaming
\end{IEEEkeywords}

\section{Introduction \& Motivation}\label{sec:introduction}
% \todo[inline]{Intro with popularity of CPS, what they are, examples}
Cyber-Physical Systems (CPS) are systems that integrate both physical and computational capabilities~\cite{baheti2011cyber}.
They can greatly assist humans at a wide variety of tasks, with examples ranging from autonomous cars, to drone systems that aid humans in search-and-rescue missions~\cite{cleland-huang_human-machine_2023,griffor_framework_2017}. 
% Frequently, CPS are classified as so-called \textit{Systems of Systems} (SoS), which consist of several independent \textit{constituent systems} (CS) that collaborate to achieve a common goal~\cite{nielsen_systems_2015}.
CPS can also exhibit adaptive characteristics, meaning they may adjust their actions and operations in response to changing parameters or changes in the environment.
A commonly used paradigm to achieve adaptability are feedback loops, such as MAPE-K\footnote{Monitor, Analyze, Plan, and Execute with shared Knowledge. Originally described by Kephart and Chess~\textcite{kephart_vision_2003}.}. 

% \todo[inline]{Human Engagement with CPS}
In many cases, CPS also integrate human-machine interaction, or even close collaboration, enabling them to benefit from the individual strengths of humans and machines. %~\cite{cleland-huang_human-machine_2023}.
In recent work, Cleland-Huang et al.~\cite{cleland-huang_human-machine_2023} mention three different paradigms of human-machine collaboration: (1) \emph{Human-in-the-Loop (HitL)} systems, (2) \emph{Human-on-the-Loop (HotL)} systems, and (3) \emph{Human-Machine Teaming (HMT)} systems.
Further, they emphasized that cooperation between humans and machines differs in each paradigm.
While HitL systems rely on humans to make decisions at key points during runtime, HotL systems operate more autonomously under the supervision of a human, who can take control if necessary~\cite{nahavandi_trusted_2017}.
The third paradigm (HMT) envisions humans and machines forming \enquote{partnerships} with shared goals, enabling seamless collaboration to optimize task efficiency.
Examples of HMT range from semi-autonomous cars~\cite{cabrall_adaptive_2018} to collaborative assembly tasks~\cite{gervasi_eye-tracking_2025}.
To achieve HMT, McDermott et al.~\textcite{mcdermott2018human} recommend defining requirements based on the three categories of \emph{transparency}, \emph{augmenting cognition}, and \emph{coordination}.

% \section{Motivation}
Commonly, a CPS that incorporates close collaboration or teaming between humans and machines directly necessitates monitoring human actors for various purposes, including collaborative activities and safety assurance.
For example, monitoring a human's position, movement, or breath data can enable the CPS to detect potential accidents and dispatch emergency responders in a smart factory~\cite{nwakanma_detection_2021}. 
Naturally, collecting sensitive data of humans raises numerous concerns about the privacy and ethics of the CPS.
HMT in CPS offers significant potential, but misuse, abuse, or defective design can lead to catastrophic events with potential human casualties~\cite{national_transportation_safety_board_collision_2020}.
Consequently, ethics and human values become a key concern and need to be considered from the design phase until the maintenance of adaptive CPS that operate within society~\cite{trentesaux_ethical_2020}. In line with this, the IEEE recommends that the design, development, and implementation of intelligent and autonomous systems should follow the principles of (1) non infringement of \emph{Human Rights}, (2) \emph{Well-being}, (3) \emph{Accountability} of designers and operators, (4) \emph{Transparency}, and (5) \emph{Awareness of misuse}~\cite{the_ieee_global_initiative_on_ethics_of_autonomous_and_intelligent_systems_ethically_2018}.

\begin{figure*}[ht]
    \centering
    \includegraphics[width=1\linewidth]{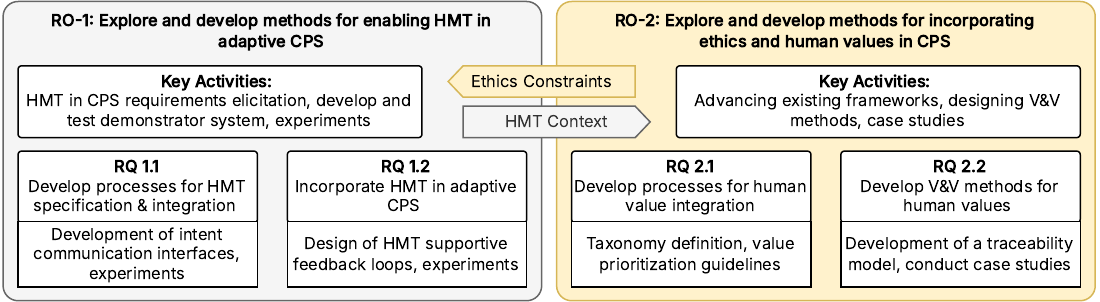}
    \caption{Graphical Overview of our Research Objectives.}
    \label{fig:research-plan}
    \vspace{-0.8em}
\end{figure*}

% In this thesis, we address the research problem 
% elicitation of requirements for effective HMT in CPS
% design and implementation of CPS with feedback loops that allow for HMT while also keeping CPS adaptability
% ensuring that requirements and design of the system conform to ethics and human values.
While the potential benefits of HMT in adaptive CPS are significant, realizing effective and ethical HMT in adaptive CPS still presents considerable challenges.
Achieving seamless teaming requires addressing complex issues such as bidirectional human-machine intent communication, accommodating varying operator knowledge, and differing operational cadences between humans and machines, all while retaining the adaptive feedback loops of CPS~\cite{cleland-huang_human-machine_2023, hoffman_inferring_2024}.
Further, the monitoring of human actors needed to support HMT raises critical privacy and ethical concerns that must be considered throughout the system lifecycle.
Existing recommendations for defining HMT requirements~\cite{mcdermott2018human} lack clear processes to specify requirements that enable both effective \emph{and} ethical HMT in CPS. 
Pertaining to the aspect of self-adaptation, there is a need to expand on, implement, and evaluate existing concepts to monitor human cognitive load in real-time, such that a machine partner can adapt to the current situation of a human~\cite{cleland-huang_human-machine_2023}.
Addressing these research problems requires creating novel or advancing existing methods for eliciting HMT requirements, designing adaptive feedback loops that incorporate teaming, and ensuring conformance with human values throughout the system lifecycle.
% my motivation to research this is: i want to see humans and machines cooperate to alleviate stress from humans, which also includes respecting human values and keeping human risk of injury at a minimum. Teaming with a robot should be fun.

\begin{figure*}
    \centering
    \includegraphics[width=.96\linewidth]{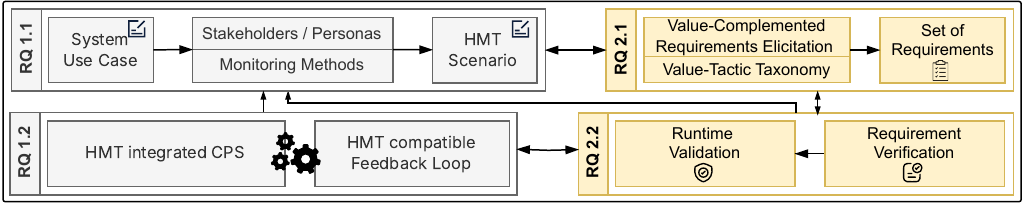}
     \vspace{-0.8em}
    \caption{Graphical Depiction of the Relationships between our Research Objectives and Planned Contributions.}
    \label{fig:research-process}
    \vspace{-0.8em}
\end{figure*}

To address these research problems, this thesis will focus on two main research objectives (cf. Figure~\ref{fig:research-plan} and~\ref{fig:research-process}): 
\begin{enumerate}[
    align=left, % Aligns the label to the left within its allocated space
    leftmargin=!,
    labelwidth=\widthof{\textbf{RO-1}}, % Sets the width of the label box to fit your widest label
]
    \item[\textbf{RO-1}] Investigate and develop novel methods for \textbf{facilitating effective Human-Machine Teaming} in adaptive Cyber-Physical Systems, and
    \item[\textbf{RO-2}] Explore and develop methods for \textbf{incorporating ethical principles and human values} into the engineering of adaptive Cyber-Physical Systems.
\end{enumerate} 
 Focusing on the first objective, our research goal is to \emph{explore current and develop novel methods of enabling effective HMT in adaptive CPS}.
As part of this, we will investigate the following research questions:

\vspace{.2em}
\noindent \textbf{RQ~1.1}: \emph{What novel methods and processes need to be developed to support the specification and integration of HMT within adaptive CPS?}\vspace{0.5em} % Process Teil / requirements / hri

Answering this question includes investigating how requirement engineering processes can be adapted to capture and document the needs and expectations for HMT in CPS, such as different operating cadences of humans and machines~\cite{cleland-huang_human-machine_2023}.
Since HMT inherently involves interaction between humans and machines, we will also consider principles from the area of Human-Robot Interaction (HRI) at this stage.
Further, we will consider bidirectional human-machine intent communication, which is vital for effective HMT. 
% how can we enrich the requirements engineering process with information required for effective human machine teaming?
For example, a transport robot must be aware of humans that may cross its path to predict and prevent possible collisions~\cite{hoffman_inferring_2024}.
With the additional information gained from HRI methods and human-machine intent communication, we strive to enrich the requirements engineering process for effective HMT, potentially by developing novel use case modeling techniques tailored for capturing dynamic HRI and intent communication in adaptive CPS.

\vspace{.2em}
\noindent \textbf{RQ~1.2}: \emph{How can we incorporate HMT in adaptive CPS while preserving the CPS's self-adaptive nature?}
\vspace{.2em}

To allow effective HMT in self-adaptive systems, we need to account for factors like varying operator expertise, trust levels between operators and machines, and differing operational cadences~\cite{cleland-huang_human-machine_2023,gil_designing_2019}.
Addressing these factors necessitates the design of self-adaptive feedback loops that explicitly integrate HMT capabilities. % add why this is important
The requirement engineering methods and human-machine communication techniques explored in RQ~1.1 serve as a foundation for developing such adaptive system capabilities.

To achieve meaningful teaming and enable adaptation, HMT involves the collection of potentially sensitive information about a human operator's actions and behavior.
This leads to ramifications that can hamper the successful integration of HMT in CPS. Therefore, our second research goal is to \emph{develop methods for embedding and ensuring adherence to human values and ethical principles during the design and operation of adaptive CPS that facilitate HMT}. As part of this, we will focus on the following research questions:

\vspace{.2em}
\noindent \textbf{RQ~2.1}: \emph{What novel requirements engineering methods are needed to enable the effective integration of human values in adaptive CPS that facilitate HMT?}\vspace{0.2em}

To avoid involuntary data sharing or misuse in HMT, human values must be integrated starting from the requirements engineering process.
In our initial work in this area~\cite{pfisterValueComplemented2025final}, we conceptualized a value-complemented framework with the goal to actively incorporate human values in the requirements engineering process of use cases involving human monitoring.
This will serve as the starting point for our research into CPS with HMT that act with human values at their core.
Our goal, thereby, is to develop novel methods to enable the integration of human values and ethics across the entire software development life cycle.
% SDLC hier erwähnen

\vspace{.2em}
\noindent \textbf{RQ~2.2}: \emph{How to verify and validate that specified human values and ethical considerations are embedded during system runtime?}\vspace{0.25em} %validation / case studies etc / 

This research question complements RQ~2.1 as it aims to verify and validate~\cite{noauthor_ieee_2017} the resulting human value integrated requirements and implementations.
This ensures that human values and privacy are followed not just throughout the design phase, but also during runtime.

% \vspace{.25em}

As our systematic literature review (SLR)~\cite{kitchenham2007guidelines} progresses, we will refine our research objectives and questions based on identified literature, potentially building upon existing methods rather than developing entirely novel ones where applicable.

% we believe that current research (such as https://doi.org/10.1109/RE48521.2020.00030) does not provide extensive end-to-end value embedded software engineering solutions.

% \todo[inline]{The research problem to be resolved, its significance, and the \textbf{justification that prior research has not yet resolved the issue}.}
% \todo[inline]{add concrete research questions, add focus of RE (why is this work important for RE?)}

\section{Related Work}\label{sec:related-work}
Human-machine collaboration is a key area of research in CPS. Recently, Cleland-Huang et al. \textcite{cleland-huang_human-machine_2023} described an approach to augment self-adaptive CPS following the \mbox{MAPE-K} feedback loop to \enquote{support partnerships between humans and machines}.
Their approach describes an evolution of HotL, where humans and machines interact more closely with one another to make collaborative decisions in a given mission.
This requires solving several problems, such as effective communication between humans and machines.
For example, the \enquote{machine} must monitor the human operator's current state and cognitive workload to adapt what information it presents to them, while the human must be aware of the status of each constituent of the CPS. 
Hoffman et al.~\cite{hoffman_inferring_2024} provide an example of close human-machine collaboration, where humans and robots assemble an object together. Here, the robot predicts the next action or infers potential positional changes, while the human provides corrections to the robot's activity.

Disregarding the human operator's workload, fatigue, and frustration may result in a higher risk of injury or lower system acceptance~\cite{hoffman_inferring_2024}. 
Morando et al.~\cite{morando_visual_2021} collected data from autonomous vehicle owners that showed that driver attention on the road and steering wheel control decreased immediately after engaging autonomous driving mode.
A human operator's failure to detect system malfunctions or lane drifts can lead to serious safety concerns or even fatal crashes.
This highlights the importance of monitoring human state to develop strategies for maintaining engagement and facilitating effective HMT. % situational awareness
Li et al.~\cite{li_hey_2021} propose a framework that provides \emph{preparatory notifications} to a human operator. This allows human operators to shift their attention to a forthcoming task, allowing for more effective task execution.
% \zp{TODO: HMT Papers}

Ethics are a major concern when working with intelligent and autonomous systems, such as CPS. 
The IEEE has released the document \emph{Ethically Aligned Design}~\cite{the_ieee_global_initiative_on_ethics_of_autonomous_and_intelligent_systems_ethically_2018} that provides a reference for integrating ethical considerations into the design and construction of these systems.
Besides these recommendations, recent studies on machine ethics in CPS~\cite{trentesaux_ethical_2020,trentesaux_engineering_2022} provide a starting point for ethical frameworks that can be considered when designing, constructing, and operating CPS.

Integrating ethics, or more generally, human values, into software engineering requires a taxonomy, such as Schwartz's theory of human values~\cite{schwartz_universals_1992}. It lists 10 core universal values and their relationships (e.g., Security opposing Self-Direction).
Whittle et al.~\cite{whittle_case_2021} applied Schwartz's taxonomy to enhance requirements engineering, using identified values to extract new requirements. They found this approach captures the \emph{why} of requirements, complementing the traditional \emph{what} related to functional and \emph{how} related to non-functional requirements.

\section{Challenges \& Anticipated Contributions}\label{sec:challenges-and-contributions}
% \todo[inline]{The research goal or hypothesis.}

% Human Monitoring at runtime is challenging because of the privacy ramifications -> values, etc. (REFSQ25) additional content at RE?

% Adaptive CPS with Human Collaboration is challenging because also cognitive overload, architectural decisions, process implementation 

% Human Machine Interaction is challenging because of cognitive overload, interface problems, etc. (see discussion with pascal)

%% 1-2 sentences between headings
In the following, we will discuss our research contributions related to the two defined research objectives (cf. Section~\ref{sec:introduction}). An overview of the relationships between our research objectives and planned contributions is shown in Figure~\ref{fig:research-process}.

\subsection{Human-Machine Teaming in adaptive CPS}\label{subsec:hmt-in-adaptive-cps}\label{subsec:HMT-in-CPS}
% \todo[inline]{Human-Machine Teaming, data collection (human monitoring), machine to human intent communication}

We consider two core challenges in HMT~\textcite{cleland-huang_human-machine_2023} as a starting point for integrating HMT in adaptive CPS (cf. RO-1): (1) designing self-adaptive MAPE-K loops that integrate HMT, considering the differing operational cadences of humans and machines, and (2) the development of methods for human-machine consensus and conflict resolution.

% To support HMT in CPS, we require humans and machines to be able to interact with one another. 
Supporting HMT in CPS requires effective human-machine interaction.
One particular area of interest is bidirectional human-machine intent communication and inference~\cite{hoffman_inferring_2024}.
A key aspect in human intent recognition is to keep cognitive load for the human operator at an appropriate level, considering different factors such as their capabilities, fatigue, or stress. %, or differences in operating cadences between machines and humans. 
Monitoring cognitive load enables robots to leverage human capabilities for more efficient collaboration and track fatigue or stress to mitigate the risk of injuries~\cite{cleland-huang_human-machine_2023,hoffman_inferring_2024}.
% By monitoring cognitive load, robots can leverage knowledge of a human's capabilities to collaborate more efficiently, while tracking fatigue or stress levels helps to reduce the risk of injuries~\cite{cleland-huang_human-machine_2023,hoffman_inferring_2024}.

We will conduct an experiment presenting human controllers with different implementations of a VR interface for receiving the intent of a robot (e.g., a TurtleBot~\cite{open_source_robotics_foundation_inc_turtlebot_nodate}) during a mission. 
With this experiment, we aim to investigate how different levels of machine intent information, presented via the VR interface, impact human operator cognitive load and performance during a simulated CPS mission involving HMT. 
% The implementations will show varying amounts of details of the Turtlebot's intent to the human controllers, such as its current action, next planned action, or long-term goal. 
While the details of measuring human cognitive load within our experiment will be informed by our SLR~\cite{kitchenham2007guidelines}, previous research used metrics like heart rate variability~\cite{lagomarsino_robot_2022} or the NASA-TLX questionnaire~\cite{hart_development_1988}. 
% Additionally, we can measure mission performance using variables such as completion time and error rate.
The results from these experiments will inform the \emph{development of principles for effective human-machine interfaces that balance intent communication with cognitive load} (cf. RQ~1.1). 
Further, understanding of the human-machine interface requirements will provide a basis for developing novel methods to \emph{incorporate HMT in adaptive CPS}, while maintaining the operator's situational awareness (cf. RQ~1.2). To support this process, we intend to map requirements to the \enquote{demons of situational awareness}~\cite{endsley2016SADemonsEnemies}, such as \emph{data overload} or \emph{complexity creep}, for enhanced traceability and additional context.
% contributing to our broader research on novel methods for HMT integration (RQ~1.1) and incorporating HMT in adaptive CPS (RQ~1.2)
 % Implementation of novel interfaces for human-machine intent communication (RQ1.1/1.2)
% goal: understand what amount of information induces cognitive stress to a critical point (i.e., could risk errors in the mission)

Monitoring features, such as human fatigue levels, necessitate the collection of sensitive data, like heart rate. 
In line with our ethical considerations (cf. RO-2), \emph{we will develop methods that enable effective human monitoring while preserving privacy}. This will involve incorporating our human-value complemented design strategies into CPS, starting from the requirements engineering process to find a balance between system requirements and privacy concerns.
To address the challenge of human-machine consensus formation~\cite{cleland-huang_human-machine_2023} during a mission, we intend to extend our existing framework~\cite{pfisterValueComplemented2025final} to include definitions of how humans and machines must act in specific situations. An interesting integration strategy for this concern could be the definition and use of a Controlled Natural Language (CNL)~\cite{gao_controlled_2016}.
This will then serve as a basis for conflict resolution during a mission (cf. RQ~1.1, 2.1).
% \zp{structured language for defining teaming / ethics attributes (CNL/DSL)}

Both of these research topics benefit from additional evaluation in the real world, i.e., in industry. We are actively looking to collaborate with industry partners during this thesis.

\subsection{Incorporating Ethics and Human Values in adaptive CPS}\label{subsec:human-values-in-CPS}
% \todo[inline]{processes of ethical requirements engineering, refsq start}
Recently, Whittle et al.~\cite{whittle_case_2021} argued that \enquote{\emph{human values are heavily underrepresented in SE methods}}.
We argue that this is particularly true when a system collects sensitive data about humans, which could potentially be misused.
A prominent example of data misuse comes from Amazon, which monitored workers in their warehouses to track and assess their performance~\cite{lecher_how_2019}.
In my thesis, we plan to develop novel methods to integrate human values and ethics into the requirements engineering process, ensuring that the specific values and ethics of different stakeholders (e.g., workers and managers) are reflected (cf. RQ~2.1).

As a starting point, we derived an initial conceptual value-complemented framework to incorporate human values as first-class citizens in the early stages of the requirements engineering process, guiding subsequent elicitation of requirements, and ultimately system design~\cite{pfisterValueComplemented2025final}.
The framework focuses on eliciting requirements for human-monitoring specific use cases.
It is based on the value taxonomy introduced by Schwartz~\cite{schwartz_universals_1992} and the work of Whittle et al.~\cite{whittle_case_2021}.

Continuing this line of research, our future work focuses on several key areas.
First, we plan to develop methods for the \emph{continuous validation and verification of value-complemented requirements} throughout the system lifecycle, including a Traceability Information Model to facilitate \mbox{(semi-)automated} checks as systems evolve~\cite{mader_getting_2009}. We plan to evaluate these concepts through participatory design studies with stakeholders (cf. RQ~2.2).
Second, we aim to create a \emph{taxonomy of monitoring value tactics}~\cite{wohlrab_supporting_2024}, based on Schwartz's taxonomy~\cite{schwartz_universals_1992}, to assist in eliciting detailed requirements for specific monitoring use cases. We aim to conduct empirical studies with companies to identify common monitoring value tactics that stakeholders deem relevant (cf. RQ~2.1).
Third, we will \emph{extend our current framework}~\cite{pfisterValueComplemented2025final} to address ethical concerns, such as how a CPS should act in high-risk situations. Further, we intend to develop processes to assist stakeholders in prioritizing their specified values. We will evaluate the framework through validation studies in real-world contexts (cf. RQ~1.1, 2.1).%, such as robotic and drone applications, exploring human-monitoring requirements (cf. RQ~1.1, 2.1).

\section{Anticipated Timeline}\label{sec:timeline}
% \todo[inline]{A timeline planned (or followed)}
% We started working on our dissertation in May 2024. % It is planned to be completed in May 2028. 
In the first year of my PhD studies (starting May 2024), we started exploring literature in the area of Systems of Systems, CPS, human-machine interaction, and ethics in computer science, which will be expanded to an SLR~\cite{kitchenham2007guidelines} in the second year.
Further, we published a first paper outlining our ideas for value-complemented requirements engineering (cf. RQ~2.1)~\cite{pfisterValueComplemented2025final}. 
With the SLR, we aim to further explore recent work related to our overall research objectives, as discussed in Section~\ref{sec:introduction}. 
Additionally, we are planning to start work on eliciting and documenting requirements needed to reduce cognitive load in HMT (cf. Section~\ref{subsec:hmt-in-adaptive-cps}).%, with the aim of publishing a paper at CHASE or MUM. 
% We aim to publish work related to integrating HMT in \emph{adaptive} CPS at SEAMS (cf. RQ~1.2).

In the later stages of the second year and the beginning of the third year, we plan to expand on our concept of value complemented software engineering~\cite{pfisterValueComplemented2025final}, making contributions to integration, validation, and verification of values throughout the software development life cycle (cf. RQ~2.2).
% We aim to publish these works at the RE conference in the future.
Over the final two years, we will continue pursuing our research objectives and finalize the thesis.
Potential venues for publishing our work include RE, CHASE, SEAMS, or MUM (International Conference
on Mobile and Ubiquitous Multimedia).
% In the remaining time of years three and four, we plan to continue working towards achieving our research objectives, as well as finalizing the PhD thesis.
% \zp{conferences hier zusammenfassen: chase, mum (HIT), SEAMS, re, etc.}

\section{Conclusion}\label{sec:conclusion}
In this research, we address current challenges of \emph{incorporating Human-Machine Teaming in adaptive Cyber-Physical Systems} while \emph{respecting the ethics and values of operators and stakeholders}.
To do so, we will investigate how requirement engineering processes can be adapted to capture and document the needs and expectations for HMT in CPS, such as effective bidirectional intent communication.
Further, we will advance research integrating HMT capabilities in the feedback loops of CPS, focusing on aspects such as operator expertise and differing operational cadences between humans and machines.
Since HMT involves the collection of sensitive information about humans, we will introduce frameworks that integrate human values as first-class citizens across the software development life cycle.
This will be accompanied by verification and validation processes, that ensure that human values are embedded throughout system design and runtime.
Our evaluation strategies include experiments and case studies.
We aim to collaborate with industry partners to evaluate our approaches in real-world contexts.
Collectively, these efforts provide novel methods and frameworks towards the realization of adaptive CPS with effective and ethical HMT.
% this research addresses the challenges ...    
% presented / focus on two research goals
% anticipated timeline
% \section*{Acknowledgements}
% I would like to thank my supervisors, Michael Vierhauser and Ruth Breu, for their continued advice and support.

% \newpage
% \balance 
\bibliographystyle{IEEEtran}
\bibliography{references}

\end{document}